\documentclass{PoS}
\usepackage{subfigure}
\usepackage{amsmath}
\usepackage{url}
\title{Investigating the Sharpe-Singleton scenario on the lattice by direct eigenvalue computation}

\ShortTitle{Investigating the Sharpe-Singleton scenario on the lattice by direct eigenvalue computation}

\author{\speaker{Joni M. Suorsa}\\
		Department of Physics and Helsinki Institute of Physics, University of Helsinki, \\P.O. Box 64, FI-00014 Helsinki, Finland\\
        E-mail: \email{joni.suorsa@helsinki.fi}}
\author{T. Rantalaiho\\
		Department of Physics and Helsinki Institute of Physics, University of Helsinki, \\P.O. Box 64, FI-00014 Helsinki, Finland\\
        E-mail: \email{teemu.rantalaiho@helsinki.fi}}
\author{K. Rummukainen\\
		Department of Physics and Helsinki Institute of Physics, University of Helsinki, \\P.O. Box 64, FI-00014 Helsinki, Finland\\
        E-mail: \email{kari.rummukainen@helsinki.fi}}
\author{K. Splittorff\\
		Discovery Centre, Niels Bohr Institute, University of Copenhagen, \\Blegdamsvej 17, 2100 Copenhagen\\
        E-mail: \email{split@nbi.dk}}
\author{David J. Weir\\
		Department of Physics and Helsinki Institute of Physics, University of Helsinki, \\P.O. Box 64, FI-00014 Helsinki, Finland\\
        E-mail: \email{david.weir@helsinki.fi}}

\abstract{
We investigate the phase structure of lattice QCD with dynamical 
Wilson fermions. Wilson chiral perturbation theory predicts that 
the Aoki phase and the Sharpe-Singleton scenario manifest themselves 
in very distinct behavior of the Wilson Dirac eigenvalue spectrum.
To test this prediction we perform a direct calculation of the 
eigenvalues of the non-Hermitian Wilson Dirac operator in dynamical 
lattice simulations. Moreover, we demonstrate an unexpected quark
mass dependence 
on the shape of the eigenvalue distribution in the positive quark mass 
side.}

\FullConference{31st International Symposium on Lattice Field Theory - LATTICE 2013\\
		July 29 - August 3, 2013\\
		Mainz, Germany}

\begin{document}

\section{Introduction}
The study of the low-energy behavior of QCD is of considerable interest 
because of the open problem of spontaneous chiral symmetry breaking. 
While the low-energy regime lies outside of perturbation theory, there 
exists an effective field theory description for it, called Chiral 
Perturbation Theory. This effective theory offers a systematic way to 
compute, order by order in the quark mass and momenta, quantities of direct 
phenomenological importance. To further validate and verify predictions 
done using Chiral Perturbation Theory, one needs first 
principles calculations of QCD. At the moment this can only be done 
using the lattice formulation of QCD. Unfortunately, some lattice 
fermion formulations break chiral symmetry explicitly at nonzero 
lattice spacing. Moreover, 
the continuum limit,  where one approaches zero lattice spacing, and the 
chiral limit, where one approaches zero quark mass, do not automatically 
commute. It is therefore imperative to have 
a good handle on the interplay between the symmetry breaking 
due to the discretization effects and due to the quark mass. 
The discretization effects depend on the lattice formulation  
and here we will focus on lattice simulations with Wilson fermions. 
In this case new phase structures, known as the Aoki 
phase \cite{Aoki:1983qi} and the Sharpe-Singleton scenario 
\cite{Sharpe:1998xm}, emerge if the 
continuum limit is taken prior to the chiral limit.

Since the discretization breaks chiral symmetry these effects can be 
included in Chiral Perturbation Theory. 
This extended effective theory is known as Wilson Chiral Perturbation Theory.
To second order in the lattice spacing, three additional 
terms in the Lagrangian describe the discretization effects 
\cite{Sharpe:1998xm,Bar:2003mh,Rupak:2002sm}. Each of these 
three terms comes with its own 
low-energy constant. Somewhat surprisingly, the sign of the 
constants are determined by the Hermiticity properties of the Wilson Dirac 
operator \cite{DSV,ADSV,HS1,KSV}. The magnitude of the three new constants 
determines whether the Aoki phase or the Sharpe-Singleton 
scenario will be realized. Which scenario is realized in an actual 
lattice simulation depends on the 
detailed setup, including the coupling and clover coefficient.

The classic way to distinguish the Sharpe-Singleton scenario from the Aoki 
phase is by observing the pion mass as a function of quark mass 
\cite{Sharpe:1998xm}. In the Aoki phase the 
pion will be massless at a non-zero critical value of the quark mass, 
whereas in the Sharpe-Singleton scenario 
the pions are massive even with zero quark mass. 
To understand this lattice induced phase structure better it is most useful 
to consider the eigenvalue spectrum of the Wilson Dirac operator $D_W$: 
the second order transition into the Aoki phase 
(and hence the massless pions) occur precisely when the quark mass reaches 
the eigenvalue distribution of 
the Wilson Dirac operator, whereas the first order Sharpe-Singleton 
scenario is realized through a collective effect on the eigenvalue 
distributions induced by the quark mass \cite{KSV}. Moreover, the distance 
from the quark mass to the eigenvalue spectrum acts as the effective quark 
mass which enters the standard form of the GOR relation \cite{KSV}. 
The main purpose of this ongoing project is to test this eigenvalue 
realization of the phase structure of simulations with dynamical Wilson 
fermions. In particular,
our aim is to observe the realization of the Sharpe-Singleton
scenario through the collective behavior of the eigenvalues.
We report here 
our current results for this project.

\section{Wilson Chiral Perturbation Theory and the Sharpe-Singleton scenario}

The Sharpe-Singleton scenario is realized through a collective 
jump of the eigenvalue distribution of the Wilson Dirac operator: for a 
positive quark mass, the eigenvalues form a band on the left side of the 
quark mass, and for a negative 
quark mass, the eigenvalues lie on the right side of 
the mass. Clearly, such a 
collective effect of the eigenvalues can only be induced by the quark mass 
in the unquenched theory. 
In contrast the quark mass only has a local effect on the Wilson Dirac 
eigenvalues in the Aoki phase.

This can be seen by calculating the two flavor eigenvalue density of 
$D_W$ using mean field Wilson chiral perturbation theory \cite{KSV} 
\begin{align}\label{eq:twofielddensity}
\rho_{c,N_f = 2}^{MF}(\hat{x}, \hat{m}; \hat{a}_6, \hat{a}_8) &= 
\frac{1}{Z_2^{MF}(\hat{m}; \hat{a}_6, \hat{a}_8)} \\
&\times \bigg{\{} e^{2\hat{m}+16|\hat{a}^2_6|-4\hat{a}^2_8}
\theta(8\hat{a}^2_8 - | \hat{x} + 16|\hat{a}_6|^2|)\nonumber \\
&+ e^{-2\hat{m}+16|\hat{a}^2_6|-4\hat{a}^2_8}
\theta(8\hat{a}^2_8 - | \hat{x} - 16|\hat{a}_6|^2|)\nonumber \\
&+\theta(8(\hat{a}^2_8 + 2\hat{a}^2_6)-|\hat{m}|)
\theta\left( 8\hat{a}^2_8 - \left| \hat{x} + \frac{2|\hat{a}_6|^2 \hat{m}}{\hat{a}^2_8 - 2|\hat{a}_6|^2}\right| \right)e^{\hat{m}^2/8(\hat{a}^2_8-2|\hat{a}^2_6|)+4\hat{a}^2_8}\bigg{\}}\nonumber.
\end{align}
Here the notation is as follows:
\begin{equation}
\hat{a}_i^2 = a^2 V W_i,\; \hat{m} = mV\Sigma
\end{equation}
where $W_i$ are the new low energy constants, $a$ is the lattice spacing, 
$m$ is the quark mass, $V$ is the system volume and $\Sigma$ is the 
chiral condensate. In the mean field limit considered here these
dimensionless variables all have a magnitude much greater than unity.

\begin{figure}
\begin{center}
\subfigure{\includegraphics[trim=7.5cm 17.5cm 7.5cm 4cm, clip=true,width=0.30\textwidth]{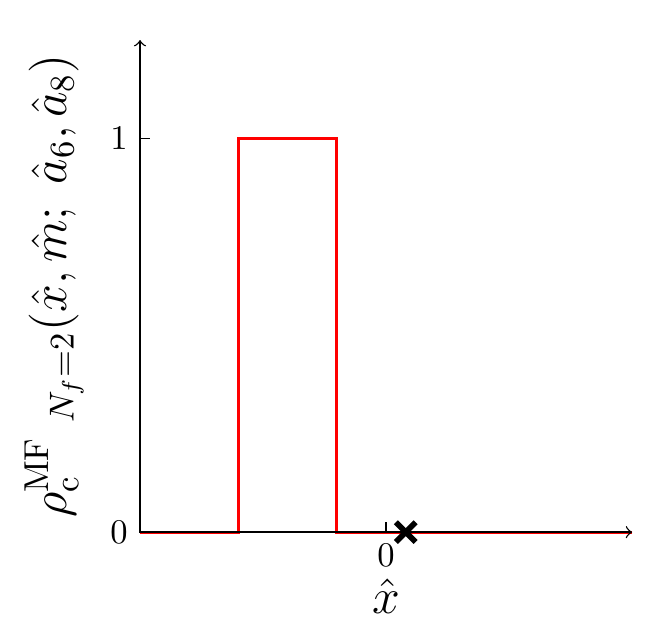}}
\hspace{0.1\textwidth}
\subfigure{\includegraphics[trim=7.5cm 17.5cm 7.5cm 4cm, clip=true,width=0.30\textwidth]{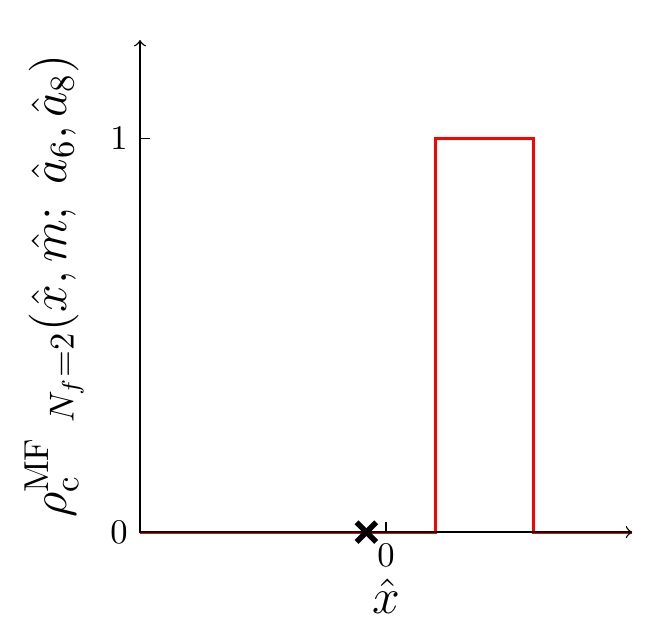}}
\end{center}
\caption{A sketch of the collective jump effect induced by a change in 
the quark mass sign in the Sharpe-Singleton scenario. The cross on the $x$-axis indicates the mass \cite{KSV}. }
\label{fig:jump}
\end{figure}

The Sharpe-Singleton scenario is realized for $\hat{a}_8^2+2\hat{a}_6^2<0$.
In this case, the effect of the quark mass changing sign 
can be seen by inspecting the contributions of the different terms in 
Eq. (\ref{eq:twofielddensity}) 
on the eigenvalue density. The first term gives rise to a strip of 
eigenvalues centered at $-16|\hat{a}^2_6|/\Sigma$ of half width of 
$8\hat{a}_8^2/\Sigma$. The second term similarly gives rise to a 
strip of eigenvalues centered at $16|\hat{a}^2_6|/\Sigma$ with the same 
half width. Both of these terms have exponential 
suppression with differing signs of mass. Now, if the mass changes sign 
from $\hat{m}$ to $-\hat{m}$, the strip of eigenvalues at $-16|\hat{a}^2_6|/\Sigma$
 jumps to $16|\hat{a}^2_6|/\Sigma$. This is sketched in 
 Fig. \ref{fig:jump}, where on the left we have a positive quark mass and 
 the distribution lies on the left of the quark mass, and on the right 
 the quark mass is negative and the distribution is on the right of the 
 quark mass.

\section{Numerical simulations}
We simulated SU(3) with two flavours on a $V = 12^4$ lattice with 
$\beta = 5.47$. Our choice of the coupling corresponded to a lattice spacing of 
$a = 0.16$ fm. 
We used Wilson fermions with clover improvement \cite{clover}, setting 
$c_{SW} = 1$, and we
implemented one level of hypercubic smearing \cite{Hasenfratz:2001hp} on the gauge fields. 
We varied the bare quark mass $m_0$ in the range $m_0 = -0.4 \hdots 0.25.$ 

We calculated fifty of the lowest lying eigenvalues of the Wilson Dirac 
operator (sorted by the real part of the eigenvalues) from one hundred 
gauge field configurations per bare quark mass choice. The eigenvalues were calculated using our 
own parallelized implementation of the Arnoldi 
algorithm which is capable of running on GPUs. The numerical complexity 
of the eigenvalue calculation for a single configuration 
was of the order of $10^{15}$ FLOPs, which warranted 
using our own implementation over the industry standard ARPACK 
\cite{ARPACK}. 

Our code implements the implicitly restarted Arnoldi method (IRAM),
with deflation \cite{Sorensen:deflation} and supports both MPI parallelization and CUDA.
Employing techniques such as kernel fusion and 
custom kernels allowed us to implement the method with minimal memory requirements
and double the performance of our initial version which was based on BLAS-like function calls. 
The parallelization framework used is based on work introduced in \cite{QCDGPU} 
and it allows us to write custom parallel kernels with complete abstraction of the
architecture. The algorithm itself is -- for the most part -- memory bandwidth bound, and 
testing our code with a $16^4$ lattice on a Tesla K20m GPU shows that we reach 
146 GB/s of effective bandwidth of the 175BG/s peak bandwidth (ECC on), meaning 
83 percent of the theoretical maximum. This makes our code on the GPU run 18.5 times
faster than ARPACK++ on one core of a Xeon X5650 @ 2.67GHz (32GB/s), measured by
the time it took for each implementation to perform the same number of Arnoldi
iterations. Scaling tests were performed using Tesla M2050 GPUs 
and the code was seen to scale well to local volumes as small as $8\times8\times16\times16$ 
(4 GPUs) with
at least a 70 percent increase for every doubling of compute resources and still a 40 percent
increase when going from 4 to 8 GPUs. Here scaling stops due to the fact that the local volume
for each GPU is too small to fill the whole processor for work.
Our implementation of the algorithm is public and free to use.\footnote{\url{
https://github.com/trantalaiho/Cuda-Arnoldi}}

We generated configurations with trajectory length of 1 for the positive 
axial quark mass configurations and calculated the eigenvalues every 
twentieth configurations. The integrator used was Omelyan 5 \cite{omelyan}. 
Similarly for the negative cases the trajectory 
length was 0.5 and the eigenvalues were calculated every 
fortieth configuration. This was because of the difficulty in simulating 
near zero axial quark mass, which required us to use conservative 
integrator time steps. Both our trajectory generation and 
the Arnoldi algorithm were run on up 
to six parallel GPUs using our GPU-capable parallel code \cite{QCDGPU}. 
Trajectory generation time ranged from around eighty 
seconds for positive quark masses to around five minutes for negative 
masses, although for half the trajectory length of the positive mass cases. 
The Arnoldi algorithm took around seven minutes per configuration.

\begin{figure}
\begin{center}
\includegraphics[trim=0.5cm 2.1cm 2.8cm 3cm,clip=true,width=0.86\textwidth]{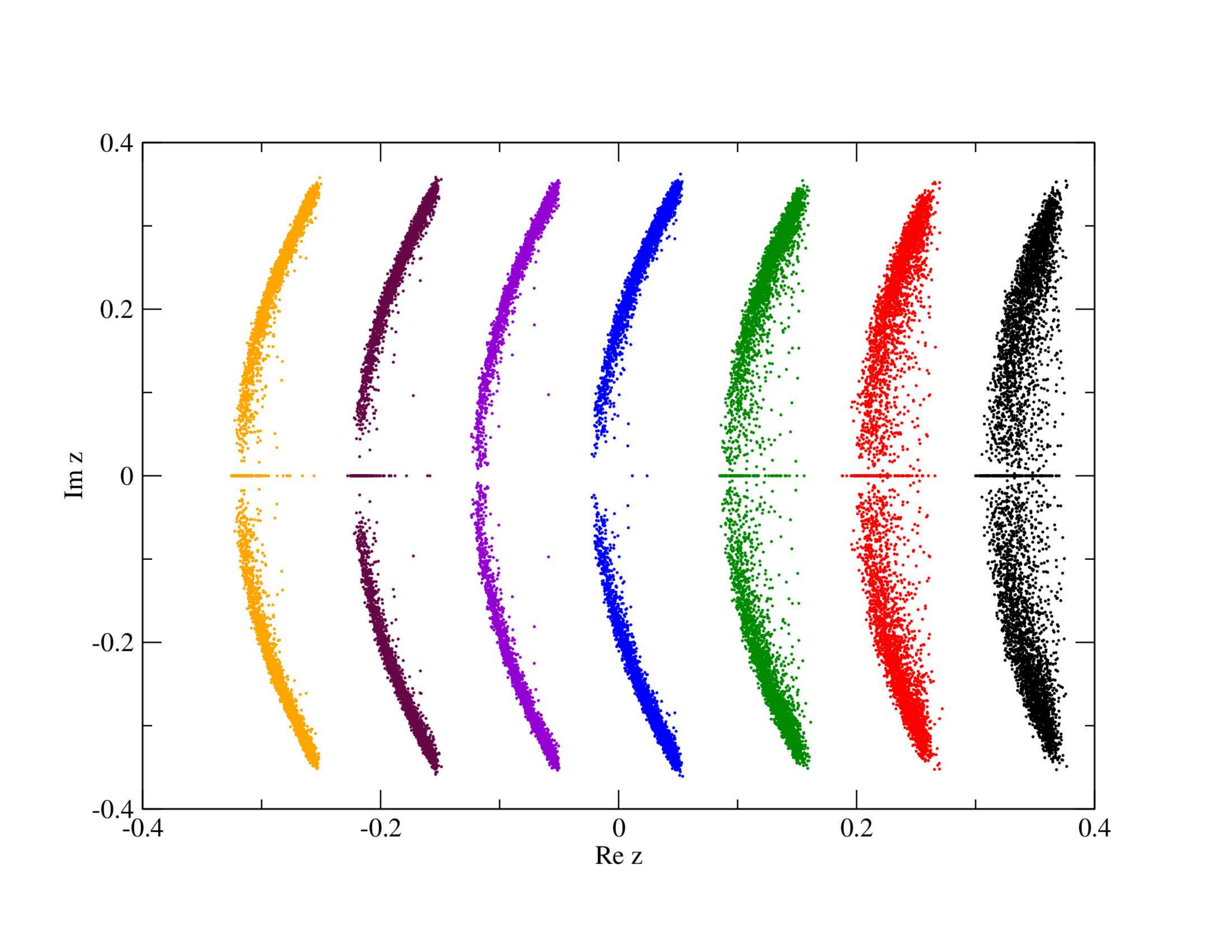}
\caption{Scatterplot of the eigenvalues $D_W + m$ for different $m_0$. From right to left the 
bare quark mass values are $0.2, 0.1, 0.0, -0.1, -0.2, -0.3$ and $-0.4.$}
\label{fig:scatterplot}
\end{center}
\end{figure}
The scatterplot in Fig. \ref{fig:scatterplot} shows the eigenvalues of $D_W+m$ 
for different values of the quark mass. We observe that with the current 
lattice setup the eigenvalue density does not show a collective jump across 
the origin. Moreover, the width of the eigenvalue distribution along the 
real axis 
as it passes zero is on the order of the smallest 
imaginary part of the eigenvalue. We conclude that the lattice effects 
in this simulation are so well under control that neither the Aoki nor the 
Sharpe-Singleton scenario is fully realized. We have computed the pion mass 
as a function of the quark mass in order to verify this. However, at the
current lattice volume these measurements are somewhat delicate. 
    
In mean field Wilson chiral 
perturbation theory to order $a^2$ the width of the eigenvalue distribution 
is independent of the quark mass, cf.~Eq.~(\ref{eq:twofielddensity}).
Unexpectedly, from this point of view, we observe a strong quark 
mass dependence in the shape of the distribution. The width 
of the distribution is seen being fairly constant in the negative 
axial quark mass side, whereas on the positive side it seems to 
grow with mass. One can also see in the figure that the zero axial 
quark mass is on the negative bare quark mass side, and that there 
is a notable depletion of real eigenvalues when one closes in on the 
zero axial quark mass.\footnote{The positive quark mass side shows a sharp right hand side edge of the 
eigenvalue distribution. This is 
not physical, but rather an effect caused by the sorting criterion of the 
eigenvalues inside the Arnoldi algorithm, where the eigenvalues were 
sorted based on the smallest real part. If one would use the smallest 
magnitude, but calculate the same number of eigenvalues, the scatterplots 
would just spread out in the real direction and shorten in the imaginary 
direction.}

\begin{figure}
\begin{center}
\includegraphics[trim=0.0cm 1.3cm 2.6cm 3cm, clip=true,width=0.45\textwidth]{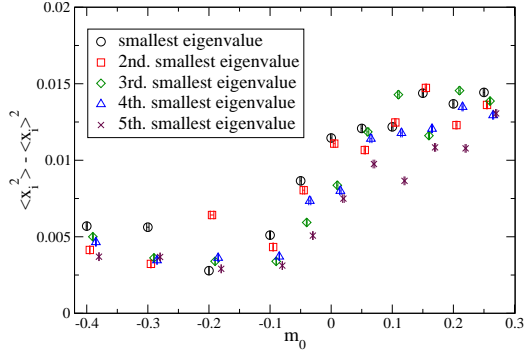}
\caption{Variance in the real axis direction for individual eigenvalues 
for various bare quark masses.}
\label{fig:variances}
\end{center}
\end{figure}

In order to quantify this, we first calculated the variance in the real 
axis direction for each individual eigenvalue. We denoted the eigenvalues 
as ordered by their imaginary parts as \\$z_i = x_i + iy_i$, with $0 < y_1 < y_2 < \hdots < y_n$
 and calculated the variance for each eigenvalue with $\sigma^2_i = \left< x_i^2 \right> - \left< x_i \right>^2$.
 The results are shown in Fig. \ref{fig:variances} for the smallest five 
of the eigenvalues calculated.

\begin{figure}
\begin{center}
\hfill
\subfigure{\includegraphics[width=0.49\textwidth]{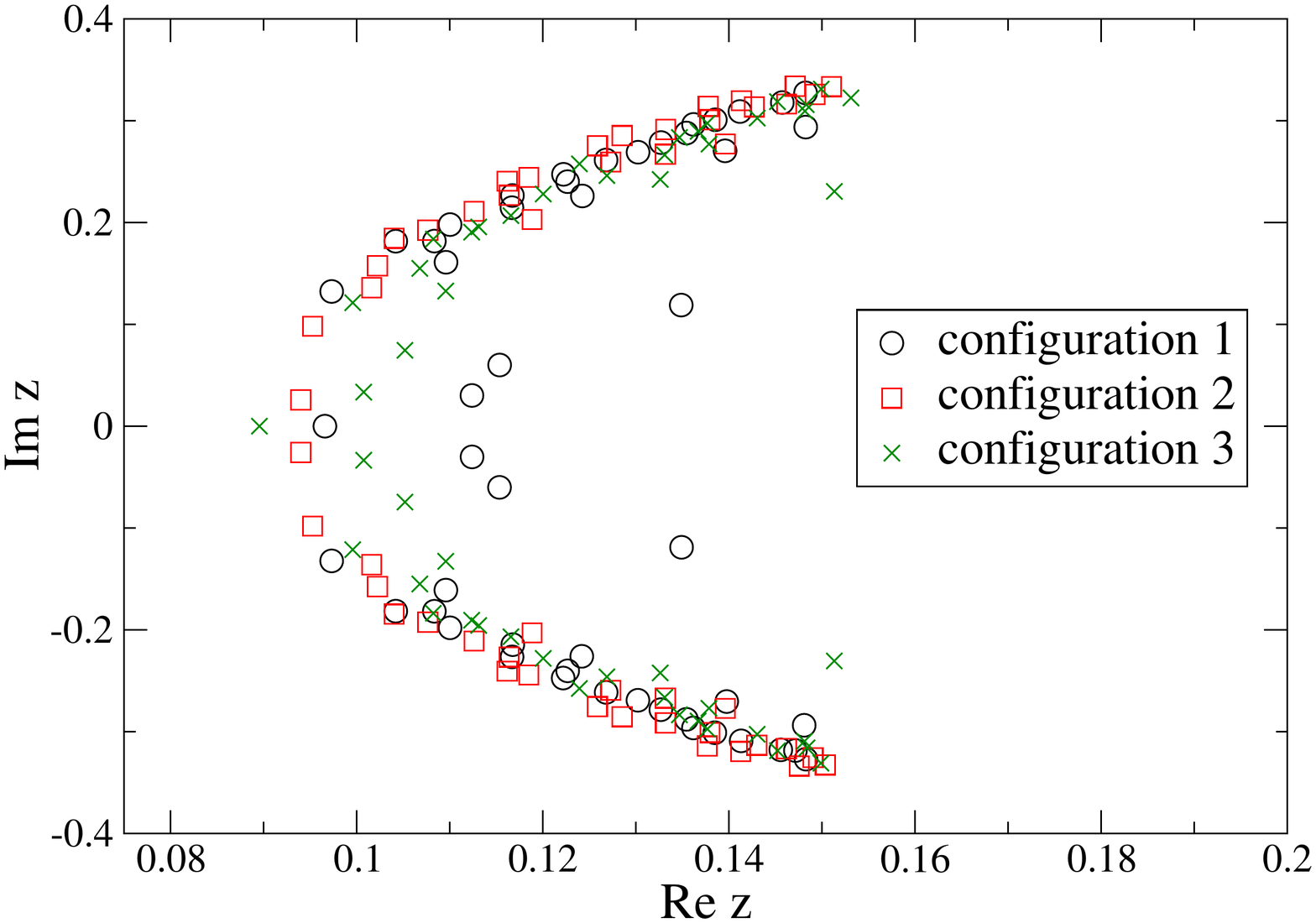}
}
\hfill
\subfigure{\includegraphics[width=0.49\textwidth]{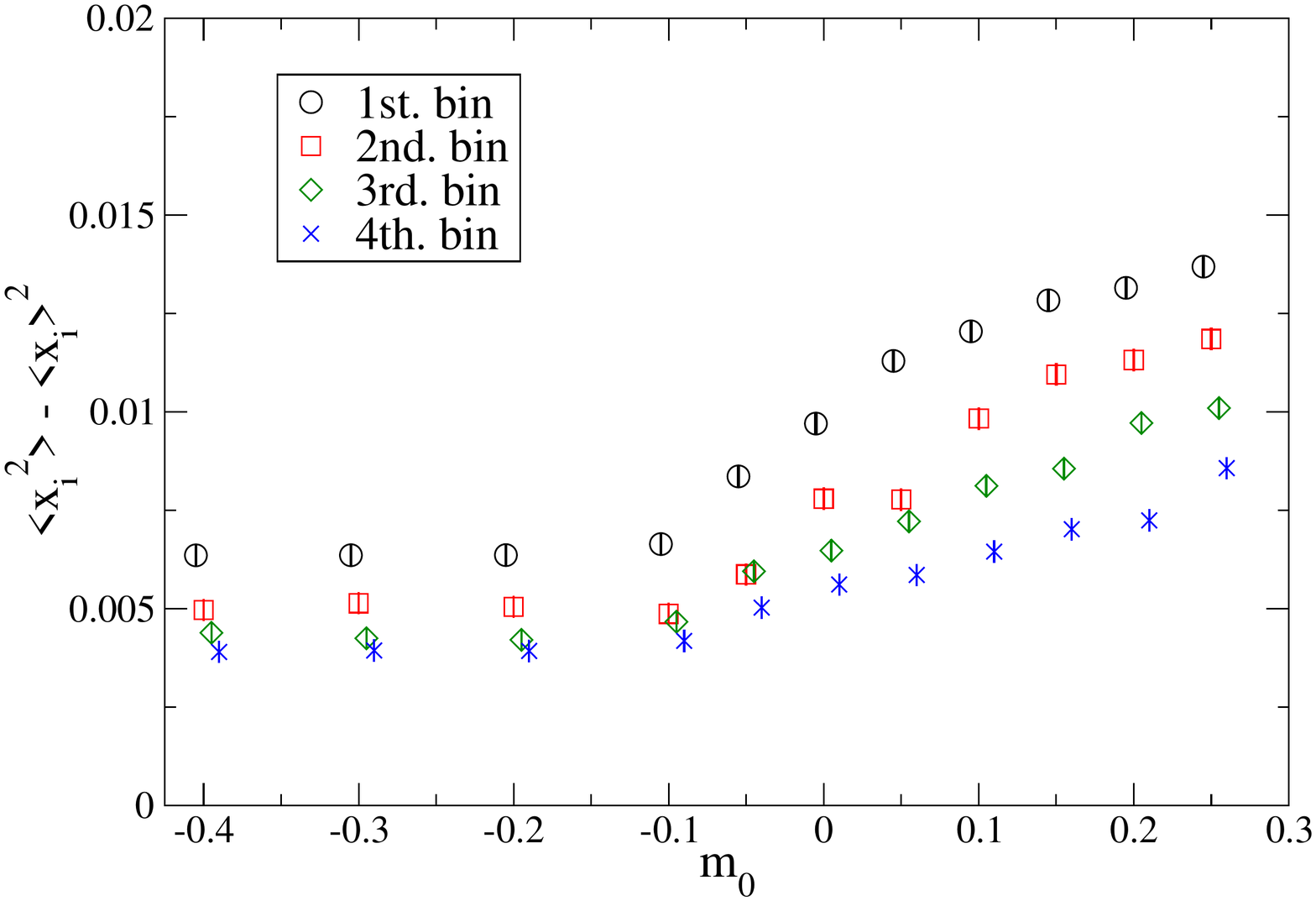}
}
\hfill
\end{center}
\caption{At left, a scatterplot of the eigenvalues calculated 
from three nonconsecutive configurations. At right, variance calculated 
by binning the eigenvalues.}
\label{fig:threeconfigsbinned}
\end{figure}

Since the data seemed quite noisy, and because of the overlap between 
nearby eigenvalues configuration by configuration, we repeated the 
variance calculation by binning the eigenvalues into groups of five.
The results of this procedure are shown in Fig. \ref{fig:threeconfigsbinned}.
It should be kept in mind that these results depend on both the number 
of eigenvalues calculated and the sorting criterion, but regardless of 
of them the mass dependence is clearly visible.

\section{Conclusions}

We are investigating the phase structure of lattice simulations with 
dynamical Wilson fermions from the perspective of the Wilson Dirac 
eigenvalues. The eigenvalues of the Wilson Dirac operator were computed 
with our own parallelized implementation of the Arnoldi algorithm.
With our present lattice setup the lattice effects are 
so well under control that the eigenvalues at small mass behave 
almost as in the continuum. Hence, from the perspective of the 
Wilson Dirac eigenvalues, the simulations does not fully enter the 
Aoki phase nor does it fully realize the Sharpe-Singleton scenario.  
We are currently computing the spectrum of the 
Hermitian Wilson Dirac operator in order to verify this. 

Our computations of the pion masses are somewhat delicate because of the 
limited lattice volume. However, with the parallelized implementation 
of the Arnoldi algorithm studies at larger volumes are also within our 
reach.

Finally, we have studied the shape of the eigenvalue 
distributions. 
Surprisingly from the point of view of Wilson chiral perturbation theory 
we demonstrated a strong mass dependence of this shape. We plan to test 
if this dependence persists in larger volumes. One possibility 
is that it is due to higher order terms in Wilson chiral perturbation 
theory, but at present no analytic predictions beyond order $a^2$ are 
available. 

\section*{Acknowledgements}
J.M.S. is supported by the Jenny and Antti Wihuri foundation. 
T.R. is supported by the Magnus Ehrnrooth foundation.
K.R. acknowledges support from Academy of Finland project 1134018.
The work of K.S. was supported by the Sapere Aude program of The
Danish Council for Independent Research.
The simulations were performed at the Finnish IT Center for Science (CSC).


\begin{thebibliography}{99}

\bibitem{Aoki:1983qi}
S.~Aoki,
Phys.\ Rev.\ D {\bf 30} (1984) 2653.

\bibitem{Sharpe:1998xm}
S.~R.~Sharpe and R.~L.~Singleton, Jr,
Phys.\ Rev.\ D {\bf 58} (1998) 074501
[hep-lat/9804028].

\bibitem{Bar:2003mh}
O.~Bar, G.~Rupak and N.~Shoresh,
Phys.\ Rev.\ D {\bf 70} (2004) 034508
[hep-lat/0306021].

\bibitem{Rupak:2002sm}
G.~Rupak and N.~Shoresh,
Phys.\ Rev.\ D {\bf 66} (2002) 054503
[hep-lat/0201019].

\bibitem{DSV} 
  P.~H.~Damgaard, K.~Splittorff and J.~J.~M.~Verbaarschot,
  Phys.\ Rev.\ Lett.\  {\bf 105} (2010) 162002
  [arXiv:1001.2937 [hep-th]].

\bibitem{ADSV} 
  G.~Akemann, P.~H.~Damgaard, K.~Splittorff and J.~J.~M.~Verbaarschot,
  Phys.\ Rev.\ D {\bf 83} (2011) 085014
  [arXiv:1012.0752 [hep-lat]].


\bibitem{HS1} 
  M.~T.~Hansen and S.~R.~Sharpe,
  Phys.\ Rev.\ D {\bf 85} (2012) 014503
  [arXiv:1111.2404 [hep-lat]].

\bibitem{HS2} 
  M.~T.~Hansen and S.~R.~Sharpe,
  Phys.\ Rev.\ D {\bf 85} (2012) 054504
  [arXiv:1112.3998 [hep-lat]].

\bibitem{KSV}
 M.~Kieburg, K.~Splittorff and J.~J.~M.~Verbaarschot,
Phys.\ Rev.\ D {\bf 85} (2012) 094011
[arXiv:1202.0620 [hep-lat]].

\bibitem{clover}
K.~Jansen and C.~Liu, 
Comput. Phys. Commun. {\bf 99} (1997) 221.

\bibitem{Hasenfratz:2001hp}
A.~Hasenfratz and F.~Knechtli,
Phys.\ Rev.\ D {\bf 64} (2001) 034504
[hep-lat/0103029].

\bibitem{ARPACK}
R.~B.~Lehoucq, D.~C.~Sorensen and C.~Yang, 
\emph{ARPACK Users' Guide} 
(SIAM, Philadelphia, 1998).

\bibitem{Sorensen:deflation}
D.~C.~Sorensen,
SIAM J. Matrix Anal. Appl. {\bf 17} (1996).


\bibitem{QCDGPU}
T.~Rantalaiho,
\emph{Porting production level quantum chromodynamics code to graphics processing units -- a case study} in \emph{Applied Parallel and Scientific Computing} (Springer, Berlin, 2013).


\bibitem{omelyan}
I.~P.~Omelyan, I.~M.~Mryglod and R.~Folk,
Comput. Phys. Commun. {\bf 151} (2003) 272.

\end{thebibliography}
\end{document}